\documentclass[%
 reprint,
 amsmath,amssymb,
 aps,
]{revtex4-2}
\usepackage{amsmath}
\usepackage{amssymb}
\usepackage[usenames]{color}
\usepackage{graphicx}% Include figure files
\usepackage{dcolumn}% Align table columns on decimal point
\usepackage{bm}% bold math
\usepackage{slashed}
\begin{document}

\preprint{APS/123-QED}

\title{Bremsstrahlung Cross Section with Polarized Beams \\for Luminosity Determination at the EIC}% Force line breaks with \\

\author{Dhevan Gangadharan}%
 \email{dhevanga@gmail.com}
\affiliation{%
 University of Houston
}%

\date{\today}

\begin{abstract}

The bremsstrahlung cross section is calculated at leading order for polarized beams of electrons and ions, which is needed for luminosity measurements at the upcoming Electron Ion Collider (EIC).  
Analytic expressions, differential in the emitted photon energy and polar angle, are derived.
The component of the cross section which depends on the beam polarizations is found to be highly suppressed with respect to the unpolarized Bethe-Heitler component, owing to the low $q^2$ that characterizes the bremsstrahlung process.

\end{abstract}

%Use showkeys class option if keyword

\maketitle

\section{\label{sec:intro}Introduction}

The dominant contribution to the inelastic cross section in lepton-nucleus collisions is the bremsstrahlung process, where a photon is emitted into the final state: $e^{-} + N \rightarrow e^{-} + N + \gamma$.
Owing to the QED nature of this process, it can be calculated to high
precision, which allowed its use to measure the collider luminosity of HERA\cite{ZEUSLuminosityGroup:2001eva,ZEUS:2013emt}.
The upcoming Electron Ion Collider (EIC) will similarly collide electrons on
protons, as well as heavier ions, and will also make use of the bremsstrahlung process to measure luminosity.
Unlike HERA, the EIC will accelerate polarized electrons \textit{and} polarized nuclei.
As the leading-order cross section only depends on the beam polarization if \textit{both} beams are polarized, the EIC requires calculation of the polarized component of the cross section.
The process has been considered before and numerically calculated for $q^2$ larger than those relevant for luminosity programs \cite{Akushevich:1998dz,Afanasev:2001jd}. 
Analytic expressions are presented here for appropriately low $q^2$.
Related calculations exist for the cases of polarized photon emission \cite{Gluckstern:1953zz} and polarized photoproduction of leptons \cite{Gehrmann:1997qh}.

\section{\label{sec:formalism}Formalism}

\begin{figure}
\includegraphics[width=0.49\textwidth]{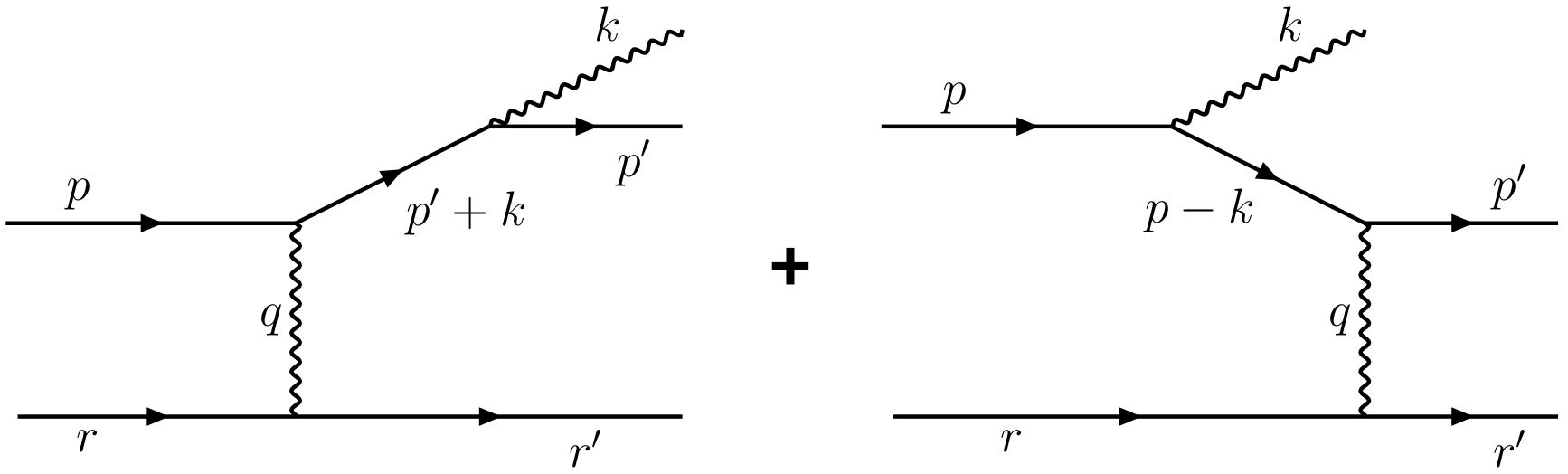}
\caption{Leading-order bremsstrahlung Feynman diagrams for $ep$ scattering.  Incoming momenta denoted by $p$ (electron) and $r$ (proton).  Outgoing momenta denoted by $p'$ (scattered electron), $r'$ (scattered proton), and $k$ (emitted photon).}
\label{fig:FeynmanDiagrams}
\end{figure}

The two Feynman diagrams that contribute to the leading-order bremsstrahlung amplitude are shown in Fig.~\ref{fig:FeynmanDiagrams}.
The corresponding unpolarized cross section was first calculated by Bethe and Heitler in 1934 \cite{Bethe:1934za}.
A more complete calculation of the cross section that goes beyond the Born
approximation (leading order) and is applicable for high atomic number $Z$, was made by Bethe and Maximon in 1954 \cite{Bethe:1954zz,Davies:1954zz,Berestetskii:1982qgu}.
Higher-order contributions enter at the subpercent level for $ep$ scattering \cite{Haas:2010bq}.
The amplitude shown in Fig.~\ref{fig:FeynmanDiagrams} is given by Eq.~\ref{eq:Amp}:
\begin{eqnarray}
M &=& e^3 \frac{ g_{\mu \nu} }{q^2} \epsilon^*_\sigma(k) \left[ \bar{u}(r') \gamma^\mu u(r) \right] \times \label{eq:Amp} \\
& & \bar{u}(p') \left[ \gamma^\sigma \frac{ \slashed{p'} + \slashed{k} + m_e }{ 2p'k } \gamma^\nu - \gamma^\nu \frac{ \slashed{p} - \slashed{k} + m_e }{ 2pk } \gamma^\sigma \right] u(p). \nonumber
\end{eqnarray}
Incoming 4-momenta (energy, momentum) in the laboratory frame are denoted as
$p=(\varepsilon,\textbf{p})$ (electron) and $r=(\varepsilon_p,\textbf{r})$
(proton).  Outgoing momenta are given by $p'=(\varepsilon',\textbf{p}')$ (scattered electron), $r'=(\varepsilon'_p,\textbf{r}')$ (scattered proton), and $k=(\omega,\textbf{k})$ (emitted photon).
The momentum of the exchanged photon is given by $q=p'+k-p$.
The electron charge is given by $e$, $m_e$ is the electron mass, $\epsilon^*(k)$ is the photon polarization vector, $u(p)$ and $u(r)$ are the spinors of the incoming electron and proton, respectively.

As in the Bethe-Heitler calculation, the same set of approximations is applied to calculate the polarized contribution to the cross section.
Since the photon polar-angle distribution of the Bethe-Heitler expression
is sharply peaked near zero, it follows that the virtuality $q^2$ of the exchange photon is also very small, $q^2 \sim m_e^2$ (Sec.~97 of Ref.~\cite{Berestetskii:1982qgu}). 
The structure of the nucleus can therefore be neglected, which is justified at low $q^2$. 
It additionally follows that this process occurs coherently with all charged nucleons in a large nucleus, and so the amplitude scales with the atomic number $Z$.
For simplicity, $ep$ scattering is considered and $Z$ is set to unity.
Owing to the smallness of the electron mass with respect to the large energies
of most experimental measurements, ultra-relativistic approximations ($p \approx \varepsilon - m_e^2/(2\varepsilon)$) are applied to the final expressions.
The ``no-recoil" approximation (Sec.~97 of Ref.~\cite{Berestetskii:1982qgu}) is applied, where the energy transferred by the exchange photon is neglected: $q = (0,\textbf{q})$.
The suitability of this approximation is discussed later in this article.

The modulus square of the amplitude in Eq.~\ref{eq:Amp} takes on the following form:
\begin{eqnarray}
|M|^2 &\equiv& -\frac{ e^6 }{ 4q^4 } W^{\mu\alpha} \, w_{\mu\alpha}, \\
W^{\mu\alpha} &=& Tr[ u(r')\bar{u}(r') \gamma^\mu u(r) \bar{u}(r) \gamma^\alpha ], \\
w_{\mu\alpha} &=& Tr[ u(p')\bar{u}(p') Q^\sigma_\mu u(p)\bar{u}(p) \bar{Q}_{\sigma \alpha} ], \\
Q^\sigma_\mu &=& \gamma^\sigma \frac{ \slashed{p'} + \slashed{k} + m_e }{ p'k } \gamma_\mu - \gamma_\mu \frac{ \slashed{p} - \slashed{k} + m_e }{ pk } \gamma^\sigma,
\end{eqnarray}
where $W^{\mu\alpha}$ is the proton tensor and $w_{\mu\alpha}$ is the electron tensor, and both are expressed as a trace over products of gamma matrices.
Typically, the final-state particle polarizations are not experimentally measurable.
Accordingly, the photon, scattered electron, and scattered proton polarizations are summed over: 
$\sum\limits_{pol} \epsilon^*_a (k) \epsilon_b (k) \rightarrow - g_{ab}$, $\sum\limits_{spin} u(p') \bar{u}(p') = \slashed{p'} + m_e$, $\sum\limits_{spin} u(r') \bar{u}(r') = \slashed{r'} + m_p$, where $m_p$ is the proton mass.
The incoming beam polarizations are measurable, for which the electron and
proton spinor products are expressed as \cite{Berestetskii:1982qgu}
\begin{eqnarray}
    u(p) \bar{u}(p) = \frac{1}{2} (\slashed{p} + m_e)(1 - \gamma^5 \slashed{a}^{(e)} ), \nonumber \\
    u(r) \bar{u}(r) = \frac{1}{2} (\slashed{r} + m_e)(1 - \gamma^5 \slashed{a}^{(p)} ). \label{eq:spinorProducts}
\end{eqnarray}
The electron and proton spin 4-vectors (Pauli–Lubanski pseudovectors) are $a^{(e)}$ and $a^{(p)}$, respectively.
They have the form $(0,\vec{\xi})$ in the particle's rest frame, for which $\vec{\xi}$ depends on the beam polarization.
For the remaining expressions, longitudinal beam polarizations are assumed.
In the target (proton) rest frame, $a^{(e)} = 2 \mathbb{P}_e \frac{ E_e \, E_p }{ m_e m_p } \left( -1, 0, 0, +1 \right)$.
The parameter $\mathbb{P}_e$ is the electron beam polarization, which is more conveniently defined as that measured in the laboratory frame along the electron's momentum.
The beam energies $E_e$ and $E_p$ are also defined as that measured in the laboratory frame.
Terms of $\mathcal{O}(m_e/\varepsilon)$ and $\mathcal{O}(m_p/\varepsilon_p)$ are neglected in $a^{(e)}$.
The proton spin 4-vector has the following form in the target rest frame, $a^{(p)} = \mathbb{P}_p  \left( 0, 0, 0, +1 \right)$.
The proton beam polarization, $\mathbb{P}_p$, is also defined as that measured in the laboratory frame along the proton's momentum.

The proton tensor, $W^{\mu\alpha}$, and the electron tensor, $w_{\mu\alpha}$, are expressed in terms of their unpolarized ($\mathcal{U}^{\mu\alpha}$, $\mathfrak{u}_{\mu\alpha}$) and polarized ($\mathcal{P}^{\mu\alpha}$, $\mathfrak{p}_{\mu\alpha}$) parts: 
\begin{eqnarray}
W^{\mu\alpha} &\equiv& \mathcal{U}^{\mu\alpha} + \mathcal{P}^{\mu\alpha}, \\
w_{\mu\alpha} &\equiv& \mathfrak{u}_{\mu\alpha} + \mathfrak{p}_{\mu\alpha},
\end{eqnarray}
where the polarized parts arise from the $\gamma^5$ terms in Eq.~\ref{eq:spinorProducts}.
Evaluating the tensor traces yields the following expressions for the polarized parts: 
\begin{eqnarray}
\mathcal{P}^{\mu\alpha} &=& 2i \, m_p \,q_a  \, a_c^{(p)} \, \varepsilon^{a \mu c \alpha}, \\
\mathfrak{p}_{\mu\alpha} &\equiv& \mathfrak{p}_{\mu\alpha}^{(1)} + \mathfrak{p}_{\mu\alpha}^{(2)} + \mathfrak{p}_{\mu\alpha}^{(3)} + \mathfrak{p}_{\mu\alpha}^{(4)}, \\
  \mathfrak{p}_{\mu\alpha}^{(1)} &=& 8i \frac{ m_e a^{(e), \lambda}}{ (p'k)^2 } \varepsilon_{a \mu \lambda \alpha} \left[ m_e^2 q^a  - p'k(p^a + k^a)\right], \\
  \mathfrak{p}_{\mu\alpha}^{(2)} &=& 8i \frac{ m_e a^{(e), \lambda}}{ (pk)^2 }  \Bigg[ (m_e^2 q^a -p'^a \, pk) \varepsilon_{a \mu \lambda \alpha} \nonumber \\ 
&& \hspace{50pt} - p'^a k_\lambda (k^b-p^b) \varepsilon_{a \mu b \alpha} \Bigg], \\
\mathfrak{p}_{\mu\alpha}^{(3)} &=& 8i \frac{ m_e a^{\lambda,(e)}}{ (pk)(p'k) }  \Bigg[ \frac{q^2}{2} p'^a \varepsilon_{\lambda a \mu \alpha} + p'^a p_\mu (k^b - p^b) \varepsilon_{\lambda a b \alpha} \nonumber \\
&& \hspace{52pt} + \bigl( p'_\alpha k^a (p^b-p'^b) \nonumber \\ 
&& \hspace{52pt} \; + p^b (p_\alpha - k_\alpha)(p'^a + k^a)\bigr)\varepsilon_{\lambda a b \mu} \Bigg], \\
\mathfrak{p}_{\mu\alpha}^{(4)} &=& -\mathfrak{p}_{\alpha\mu}^{(3)}.
\end{eqnarray}
The polarized tensors, $\mathcal{P}^{\mu\alpha}$ and $\mathfrak{p}_{\mu\alpha}$, are antisymmetric, while the unpolarized tensors, $\mathcal{U}^{\mu\alpha}$ and $\mathfrak{u}_{\mu\alpha}$, are symmetric.  
Thus, only $\mathcal{U}^{\mu\alpha} \mathfrak{u}_{\mu\alpha}$ and $\mathcal{P}^{\mu\alpha} \mathfrak{p}_{\mu\alpha}$ contribute to the cross section.
However, at higher orders with loop corrections, single-spin asymmetries emerge \cite{Afanasev:2005pb}.
We have:
\begin{eqnarray}
|M|^2 &=& -\frac{ e^6 }{ 4q^4 } \left[ \mathcal{U}^{\mu\alpha} \mathfrak{u}_{\mu\alpha} + \mathcal{P}^{\mu\alpha} \mathfrak{p}_{\mu\alpha} \right] \\
    &=& -\frac{ (4\pi)^3 \, m_e^2 \, \alpha \, r_e^2 }{ 4q^4 } \left[ \mathcal{U}^{\mu\alpha} \mathfrak{u}_{\mu\alpha} + \mathcal{P}^{\mu\alpha} \mathfrak{p}_{\mu\alpha} \right] \label{eq:M2}
\end{eqnarray}
where the fine-structure constant is given by $\alpha = \frac{e^2}{4\pi}$ and
the classical electron radius is given by $r_e =  \frac{e^2}{4\pi \, m_e}$.

\section{\label{sec:crosssections}Differential Cross Sections}

The fully differential cross section expressed in the target rest frame is
\begin{eqnarray}
    d\sigma &=& \frac{1}{(4\pi)^5} |M|^2 \frac{ |\textbf{p}'| \omega d\omega }{ |\textbf{p}| m_p^2 } d\Omega' d\Omega_k, \label{eq:DiffCS} \\
&\equiv& d\sigma_{\mathcal{U}} + d\sigma_{\mathcal{P}}.
\end{eqnarray}
Inserting the polarized part of Eq.~\ref{eq:M2} into Eq.~\ref{eq:DiffCS} gives
\begin{equation}
	d\sigma_{\mathcal{P}} = \frac{ -\alpha \, r_e^2 \, m_e^2 }{ 4(4\pi)^2 q^4 \, m_p^2 } \mathcal{P}^{\mu\alpha} \mathfrak{p}_{\mu\alpha} \frac{ |\textbf{p}'| \omega d\omega }{ |\textbf{p}| } d\Omega' d\Omega_k . \label{eq:generalPolCrossSection}
\end{equation}
The unpolarized term, $d\sigma_{\mathcal{U}}$, corresponds to the usual Bethe-Heitler expression and was re-derived as a cross check.
The angular phase space of the scattered electron and emitted photon are denoted by $d\Omega'$ and $d\Omega_k$, respectively.
In order to provide a practical expression, integration over the angles in the final state is needed.

Evaluating the contraction of Levi-Civita symbols in $\mathcal{P}^{\mu \nu} \mathfrak{p}_{\mu \nu}$ leads to the following:
{\small
\begin{eqnarray}
	\mathcal{P}^{\mu \nu} \mathfrak{p}_{\mu \nu} &=& -32 m_e m_p \Biggl[  \frac{1}{(pk)^2} \frac{q^2}{2} ka^{(e)}\Big(qa^{(p)}-2p'a^{(p)} \Big) \label{eq:fullyDif} \\ 
&+& \Big( \frac{1}{(pk)^2} + \frac{1}{(p'k)^2} \Big) \Big( qa^{(e)} qa^{(p)} m_e^2 - a^{(e)}a^{(p)} q^2 m_e^2 \Big)  \nonumber \\ 
&+& \frac{1}{(pk)(p'k)} \Big( -q^4 a^{(e)}a^{(p)} - 2qa^{(e)} qa^{(p)}m_e^2 \nonumber \\
&& \hspace{46pt} + \frac{q^2}{2}\Big( 2p'a^{(e)}( p'a^{(p)}-pa^{(p)}) \nonumber \\
&& \hspace{71pt} + ka^{(e)}(3qa^{(p)} + 2pa^{(p)}) \nonumber \\ 
&& \hspace{71pt} + 4m_e^2 a^{(e)}a^{(p)} \Big) \Big) \nonumber \\
&+& \Big( \frac{1}{p'k} - \frac{1}{pk} \Big) \Big( (qa^{(e)} + ka^{(e)})qa^{(p)} - 2q^2 a^{(e)}a^{(p)} \Big) \Biggr]. \nonumber
\end{eqnarray}
}
Integration of the differential cross section over the scattered electron
angles, $d\Omega'$, can be performed analytically \cite{Gluckstern:1953zz} and
is done first.
For longitudinally polarized beams, all 4-vector products involving the
scattered electron momentum, $p'$, in Eq.~\ref{eq:fullyDif} can be expressed in
terms of two basis vectors, $\textbf{a}$ and $\textbf{b}$, defined through
$q^2$ and $p'k$:
\begin{eqnarray}
q^2 &=& -\textbf{q}^2 = -(\textbf{p}' - \textbf{p} + \textbf{k})^2,  \nonumber \\
&\equiv& -(\textbf{p}' - \textbf{T})^2 = -(\textbf{p}'^2 + T^2)(1 - \textbf{p}'\textbf{a}), \\
p'k &=& \omega \varepsilon' - \textbf{p}'\textbf{k} \equiv \omega \varepsilon'(1 - \textbf{p}'\textbf{b}), \\
\textbf{a} &=& \frac{2\textbf{T}}{\textbf{p}'^2 + T^2}, \\
\textbf{b} &=& \frac{\textbf{k}}{\omega \varepsilon'}.
\end{eqnarray}
The other 4-vector products containing $p'$, $p'a^{(e)}$ and $p'a^{(p)}$, are also
expressed with terms containing $(1 - \textbf{p}'\textbf{a})$ and $(1
- \textbf{p}'\textbf{b})$.
Ultimately, the integral over the scattered electron angles is of the form:
\begin{equation}
I_{m,n} = \int d\Omega' (1 - \textbf{p}'\textbf{a})^{-m} (1 - \textbf{p}'\textbf{b})^{-n}.
\end{equation}
Note that by convention a factor of $1/(2\pi)$ as defined for $I_{m,n}$ in
Ref.~\cite{Gluckstern:1953zz} is not included here.
The array of integrals present is: $I_{0,0}$, $I_{0,1}$, $I_{1,0}$, $I_{-1,1}$, $I_{1,-1}$, $I_{1,1}$, $I_{2,0}$, $I_{0,2}$, $I_{2,1}$, $I_{1,2}$, $I_{2,-1}$, $I_{2,-2}$, $I_{2,2}$.
Integrals with negative indices can be expressed in terms of the other integrals.
For the non-trivial integrals, $I_{1,1}$, $I_{2,1}$, $I_{1,2}$, $I_{2,2}$, Feynman parameters are first used to combine denominators.  
A strategically chosen polar angle axis is then chosen to make the azimuthal integrations trivial.  
Finally, a table of integrals \cite{Gradshteyn:1943cpj} is used for the final integration over the Feynman parameter.
Integration over $d\Omega_k$ similarly starts with a strategically chosen polar axis ($\hat{\textbf{p}}$) that makes the azimuthal integration trivial.
Assembly of all terms resulting from each integral is quite algebraically laborious. 
Mathematica is used to assemble and simplify the algebra.

The expressions for the double- and single-differential cross sections are given in the laboratory frame.  
For the small angles that characterize the Bethe-Heitler expression at ultra-relativistic energies, the Lorentz transformation from the lab frame (LF) to the target rest frame (TRF) is especially simple: $E_{TRF} = \frac{2\epsilon_p}{m_p} E_{LF}$ for $E \in \{\varepsilon, \varepsilon', \omega\}$.
After transforming to the lab frame, the beam parameters of $a^{(e)}$ can be mapped to other variables: $E_e \rightarrow \varepsilon$ and $E_p \rightarrow \varepsilon_p$.

The resulting polarized cross section, double-differential in the photon polar angle and energy up to $\mathcal{O}(m_e^2)$, in the laboratory frame, is given by
\begin{eqnarray}
\frac{ d\sigma_{\mathcal{P}} }{ d\omega d\delta } &=& \mathbb{P}_e \mathbb{P}_p \frac{ \alpha \, r_e^2 \, m_e^2 \, \delta }{\omega \varepsilon^3 \varepsilon' \varepsilon_p (1+\delta^2)^2} \Biggl[ \delta _0^2 \Biggl(4 L_1 \varepsilon \varepsilon'^2  \nonumber \\
	&& + 2 \omega \Bigl((L_1+L_2-L_\theta) \varepsilon^2 + L_1 \varepsilon \varepsilon' - L_2 \varepsilon'^2 \Bigl) \nonumber \\ 
	&& - 2 \omega^3 - 5 \varepsilon' \omega^2 - 6 \varepsilon'^2 \omega - 4 \varepsilon'^3 \Biggr) \nonumber \\
	&& + 2\varepsilon \Biggl((1+2 L_\theta) \varepsilon'^2 + (1 - L_1 - L_2 + L_\theta) \omega^2 \nonumber \\
	&& + \varepsilon' \omega \Bigl(-L_1-L_2 + 2L_\theta + \frac{1}{1+\delta _0^2} \Bigr) \Biggr) \Biggr],
\end{eqnarray}
where $L_1 = \ln{ \frac{ 4 \, \varepsilon' \, \varepsilon_p}{ m_e m_p } }$, $L_{2} = \ln{ \frac{ 4 \, \varepsilon \, \varepsilon' \, \varepsilon_p}{ \omega \, m_e m_p } }$, $L_{\theta} = \ln{ (1 + \delta^2) }$.
One observes that, like the Bethe-Heitler spectrum, the angular spectrum is sharply peaked for $\delta = \theta_k \frac{\varepsilon}{m_e} \lesssim 1$.

\begin{figure}
\includegraphics[width=0.49\textwidth]{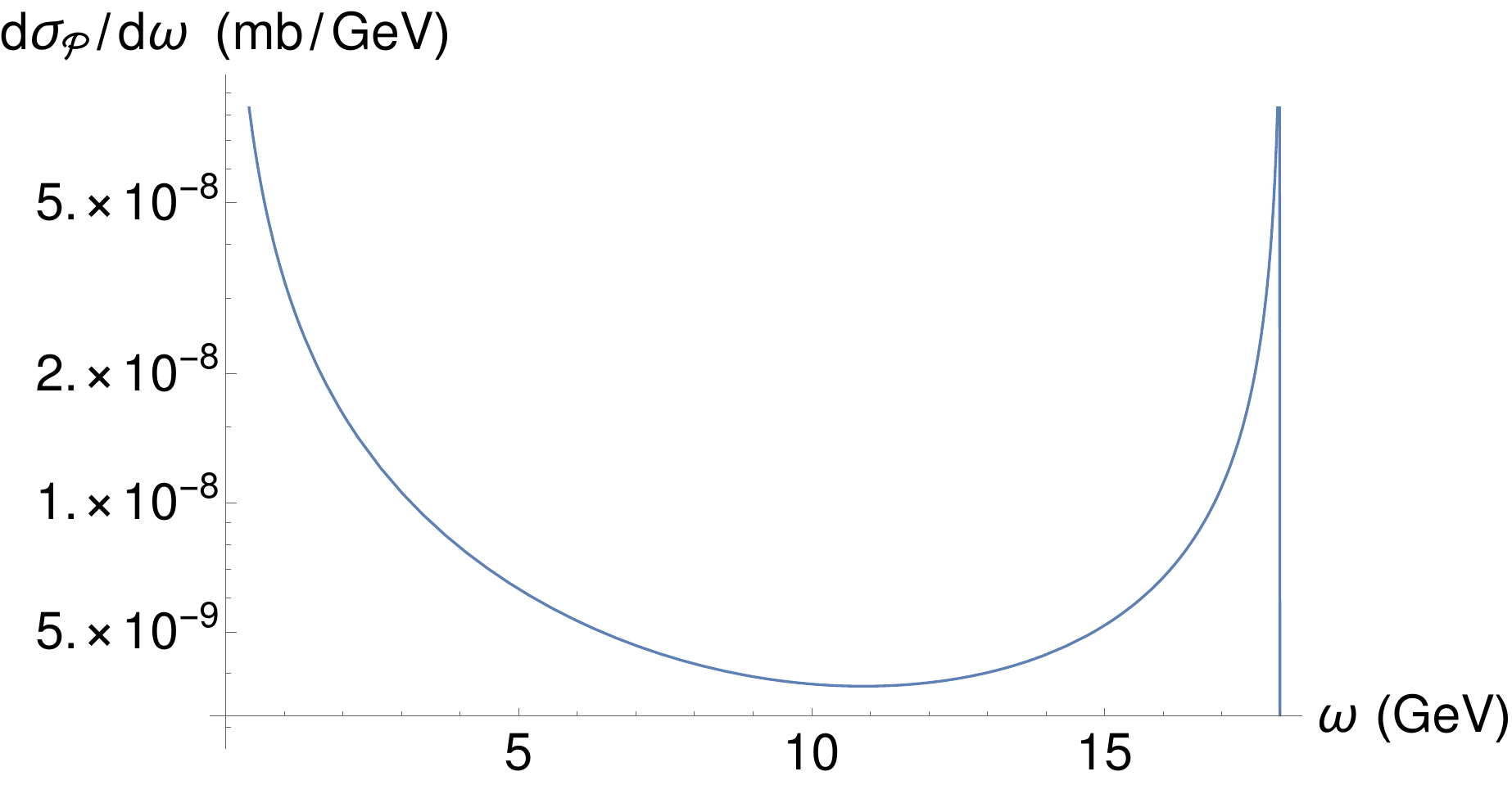}
\caption{Polarized component of the bremsstrahlung cross section versus photon energy for $\varepsilon=18$ GeV and $\varepsilon_p=275$ GeV (top EIC energies). The product of beam polarizations is set to unity: $\mathbb{P}_e \mathbb{P}_p = 1$.}
\label{fig:PBH}
\end{figure}

\begin{figure}
\includegraphics[width=0.49\textwidth]{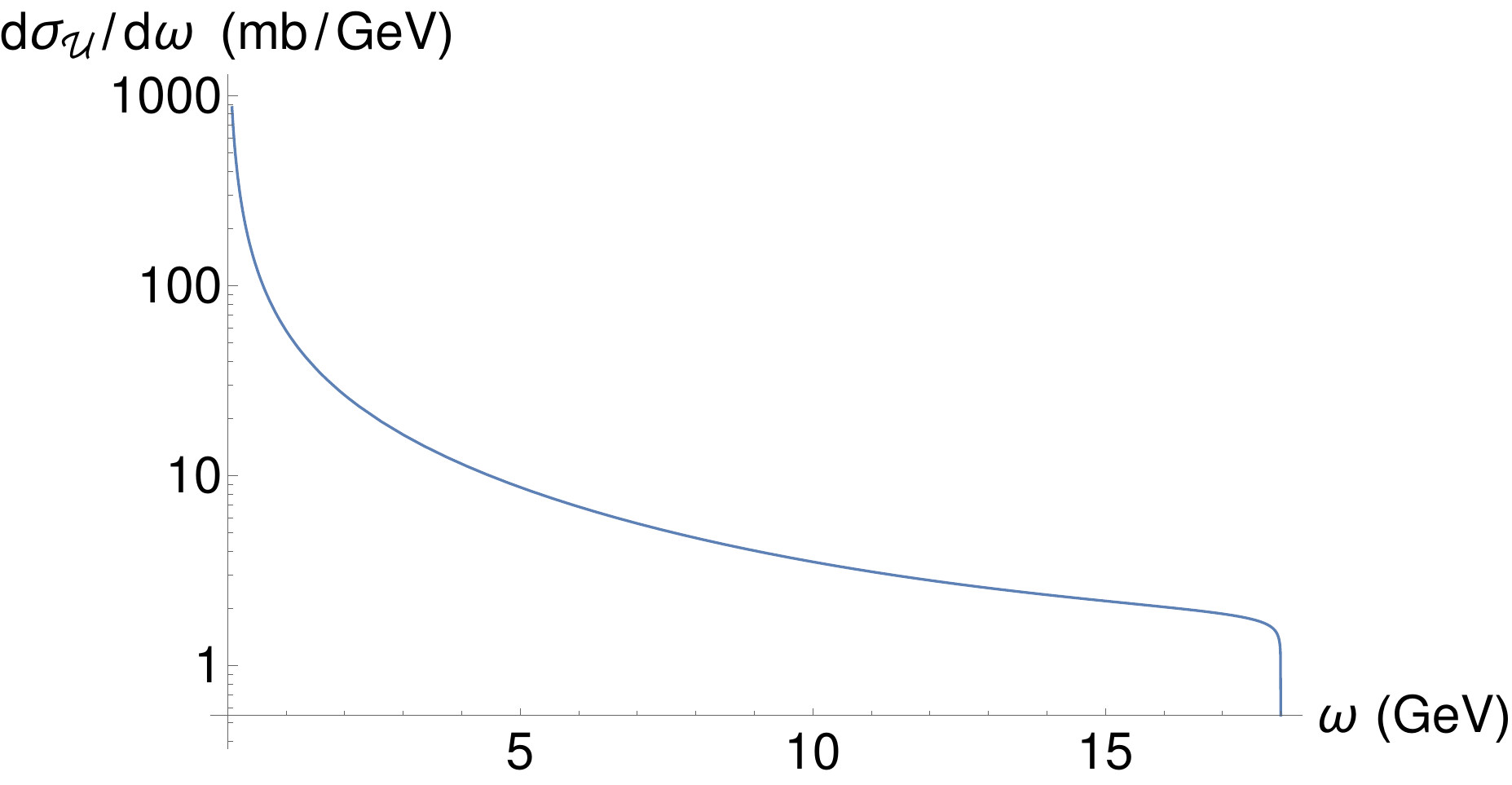}
    \caption{Unpolarized component of the bremsstrahlung cross section (Bethe-Heitler) versus photon energy for $\varepsilon=18$ GeV and $\varepsilon_p=275$ GeV (top EIC energies).}
\label{fig:UPBH}
\end{figure}

Finally, integrating over the photon polar angle yields the single-differential
polarized cross section
\begin{equation}
\frac{d\sigma_{\mathcal{P}}}{d\omega} = \mathbb{P}_e \mathbb{P}_p \frac{ 4 \alpha r_e^2 }{\omega} \frac{\varepsilon'}{\varepsilon} \frac{ m_e^2 }{ \varepsilon \, \varepsilon_p } \left( F_1 + \frac{ \varepsilon }{ 4 \varepsilon' } F_2 + \frac{ \varepsilon' }{ 8 \varepsilon } F_3 + \frac{ \varepsilon^2 }{ 2 \varepsilon'^2 } F_4 \right), \label{eq:SingleCrossSection} 
\end{equation}
where
\begin{eqnarray}
F_1 &=& \frac{1}{8} \left( 7 + L_2 (2 - 4 L_3) - 4 L_3 + L_1 (-2 + 4 L_3) \right), \nonumber \\
F_2 &=& -3 + L_1 + 2 L_2 + L_3(1 - 2 L_2 + 2 L_3), \nonumber \\
F_3 &=& (-1 + 2 L_2) (-1 + 2 L_3), \nonumber \\
F_4 &=& (-1 + L_3)(-2 + L_1 +L_2 - L_3). \nonumber
\end{eqnarray}
and $L_3=\ln{ \frac{ \pi \varepsilon \, \varepsilon_p}{ m_e m_p } }$.
The polarized cross section is to be compared to the unpolarized Bethe-Heitler expression \cite{Bethe:1934za}:
\begin{equation}
\frac{d\sigma_{\mathcal{U}}}{d\omega} = \frac{ 4 \alpha r_e^2 }{\omega} \frac{\varepsilon'}{\varepsilon} \left( \frac{ \varepsilon }{ \varepsilon' } + \frac{ \varepsilon' }{ \varepsilon } - \frac{2}{3} \right) \left[ L_2 - \frac{1}{2} \right]. \label{eq:BH} 
\end{equation}
It is clear that the polarized component of the bremsstrahlung cross section is highly suppressed with respect to the unpolarized component by a factor of $m_e^2 / (\varepsilon \, \varepsilon_p)$.
Figures \ref{fig:PBH} and \ref{fig:UPBH} show the energy spectra of the polarized and unpolarized components, respectively, for $\varepsilon=18$ GeV and $\varepsilon_p=275$ GeV (top EIC energies).
The sharp turn in the spectrum near the photon energy limit (18 GeV) in Fig.~\ref{fig:PBH} occurs when logs with different signs change in magnitude in $F_4$. 
Furthermore, the polarized component becomes negative very near to the energy limit, which can occur for perturbative calculations at finite order. 
A similar feature is also observed in a numerical calculation at larger $q^2$ \cite{Akushevich:1998dz}.
Very near to the upper energy limit, one should also keep in mind that the ultrarelativistic limit for the scattered electron breaks down.
Additionally, inclusion of the ion's small recoil energy may alter this observed feature. 

The recoil of the ion was neglected in this calculation as it was in the original Bethe-Heitler calculation, $q=(0,\textbf{q})$. 
To a better approximation, one may instead employ the relation $q=(\textbf{q}^2/2m_p, \textbf{q})$, owing to the smallness of the momentum transferred to the nucleus.  
As a consequence, terms containing $1/q^2$ in Eq.~\ref{eq:fullyDif} (inserting Eq.~\ref{eq:generalPolCrossSection}) can be expanded in a power series, for which the first two terms are $-1/\textbf{q}^2 - 1/(4m_p^2)$.
The latter extra term is suppressed by $m_e^2/m_p^2$ with respect to the first term.
Additionally, in order to simplify the algebra in this calculation, frequent use of the simple relation $\varepsilon' + \omega - \varepsilon = 0$ was used, which would have to be promoted to $\varepsilon' + \omega - \varepsilon = -\textbf{q}^2/2m_p$.
Due to the similarly small scale of the calculated polarized cross section, it is not clear how much the no-recoil approximation affects its functional form.
However, the polarized component is expected to remain suppressed.

For the case of longitudinally polarized electrons and \textit{transversely} polarized protons, as expected for the EIC, the polarized component vanishes exactly when integrating over azimuthal angles of the final-state particles.
Additionally, the vectors $\textbf{a}$ and $\textbf{b}$ do not form a complete
basis to express the 4-vector products in Eq.~\ref{eq:fullyDif}, which further
complicates analytic integration.  
Transverse polarization is not considered further here.

In regards to the general luminosity program at the EIC, where high $Z$ heavy-ions will be accelerated, it should be noted that the Born approximation underlying the Bethe-Heitler expression is known to be inadequate.
The Bethe-Heitler expression should be replaced by the Bethe-Maximon expression \cite{Bethe:1954zz,Davies:1954zz,Berestetskii:1982qgu}:
{\footnotesize
\begin{eqnarray}
\frac{d\sigma_{\mathcal{U}}}{d\omega} &=& \frac{ 4 Z^2 \alpha r_e^2 }{\omega} \frac{\varepsilon'}{\varepsilon} \left( \frac{ \varepsilon }{ \varepsilon' } + \frac{ \varepsilon' }{ \varepsilon } - \frac{2}{3} \right) \left[ L_2 - \frac{1}{2} - f(\alpha Z) \right], \label{eq:BM} \\
    f(\alpha Z) &=& (\alpha Z)^2 \sum\limits_{n=1}^{\infty} \frac{ 1 }{ n(n^2 + (\alpha Z)^2) }.
\end{eqnarray}
}
For heavy nuclei such as uranium, the Bethe-Maximon and Bethe-Heitler expressions differ by about $2\%$.
For light nuclei, $\alpha Z \ll 1$ and $f(\alpha Z) \approx 1.2 (\alpha Z)^2$.

\section{\label{sec:conclusion}Conclusion}

The leading order bremsstrahlung cross section for polarized incoming beams has been calculated in anticipation of luminosity measurements at the EIC.
An analytic expression has been derived, which shows that the polarized component is highly suppressed with respect to the usual unpolarized Bethe-Heitler expression.
The suppression is linked to the low $q^2$ that characterizes the Bethe-Heitler process, which is $\sim m_e^2$.
Further work is needed to estimate the effect of the no-recoil approximation, although the polarized component is expected to remain suppressed.

\begin{acknowledgments}
I would like to thank Andrei Afanasev, Wim Cosyn, and Katarzyna Wichmann for useful discussions.
This work is supported by US DOE Nuclear Physics Grant No.~DE-FG02-07ER41521.

\end{acknowledgments}

\bibliography{bibliography}% Produces the bibliography via BibTeX.

\end{document}